\documentclass[structabstract]{aa}  
\usepackage{graphicx}
\usepackage{txfonts}
\usepackage{natbib,twoopt}
\usepackage[breaklinks=true]{hyperref} 
\bibpunct{(}{)}{;}{a}{}{,} 
\newcommandtwoopt{\citeads}[3][][]{\href{http://adsabs.harvard.edu/abs/#3}%
{\citealp[#1][#2]{#3}}} 
\newcommandtwoopt{\citepads}[3][][]{\href{http://adsabs.harvard.edu/abs/#3}%
{\citep[#1][#2]{#3}}} 
\newcommandtwoopt{\citetads}[3][][]{\href{http://adsabs.harvard.edu/abs/#3}%
{\citet[#1][#2]{#3}}} 
\newcommandtwoopt{\citeyearads}[3][][]%
{\href{http://adsabs.harvard.edu/abs/#3}{\citeyear[#1][#2]{#3}}}
\begin{document} 

  \title{Radial velocities and metallicities from infrared \ion{Ca}{II} triplet spectroscopy of open clusters}

  \subtitle{II. Berkeley~23, King~1, NGC~559, NGC~6603, and 
NGC~7245\fnmsep\thanks{Based on observations made with the 2.5 Isaac Newton Telescope operated on the 
island of La Palma by the Isaac Newton Group in the Spanish Observatorio del Roque de los Muchachos of the Instituto de Astrof\'{\i}sica de Canarias.}}

  \author{R. Carrera\inst{1,2}
         \and
         L. Casamiquela\inst{3}
        \and
         N. Ospina\inst{4}
        \and
         L. Balaguer-N\'u\~nez\inst{3}
        \and
         C. Jordi\inst{3}
        \and
        L. Monteagudo\inst{1,2}}

  \institute{Instituto de Astrof\'{\i}sica de Canarias, La Laguna, Tenerife, Spain\\
            \email{rcarrera@iac.es}
            \and
            Departamento de Astrof\'{\i}sica, Universidad de La Laguna, Tenerife, Spain
            \and
             Departament d'Astronomia i Meteorologia, Universitat de Barcelona, ICC/IEEC, Barcelona, Spain 
             \and
             Institut de Ci\`{e}ncies de l'Espai (CSIC-IEEC), Campus UAB, Bellaterra, Spain
          }

  \date{Received September 15, 1996; accepted March 16, 1997}


 \abstract
  {Open clusters are key to studying the formation and 
evolution of the Galactic disc. However, there is a deficiency of radial 
velocity and chemical abundance determinations for open clusters in the 
literature.}
  {We intend to increase the number of determinations of 
radial velocities and metallicities from spectroscopy for open clusters.}
  {We acquired medium-resolution spectra ($R\sim 8000$) in the infrared region  \ion{Ca}{II}
triplet lines ($\sim$8500\,\AA) for several stars in five open clusters with the 
long-slit IDS  spectrograph on 
the 2.5~m Isaac Newton Telescope (Roque de los Muchachos Observatory, Spain).
Radial velocities were obtained by cross-correlation fitting techniques. 
The relationships available in the literature between the strength of 
infrared \ion{Ca}{II} lines and metallicity were also used to derive
the metallicity for each cluster.}
  {We obtain $\left\langle V_{\rm r}\right\rangle = 48.6\pm3.4$, $-58.4\pm6.8$, 
$26.0\pm4.3,$ and $-65.3\pm3.2$ km s$^{-1}$ for Berkeley~23, NGC~559, NGC~6603,
and NGC~7245, respectively. We found [Fe/H]$ =-0.25\pm0.14$ and 
$-0.15\pm0.18$ for NGC~559 and NGC~7245, respectively. Berkeley~23 has 
low metallicity, [Fe/H] $=-0.42\pm0.13$, which is similar to 
other open clusters in the outskirts of the Galactic disc. In contrast, we 
derived high metallicity ([Fe/H] $=+0.43\pm0.15$)  for NGC~6603, which places 
this system amongst the most metal-rich known open clusters. To our knowledge, 
this is the first determination of radial velocities and metallicities from 
spectroscopy for these clusters, except  NGC~6603, for which radial velocities had been 
previously determined. We have also analysed ten stars in the line of sight to
King~1. Because of the large dispersion obtained in both radial velocity and metallicity,
we cannot be sure that we have sampled true cluster members.}
{}

  \maketitle
%

\section{Introduction}

Stellar clusters are crucial to the study of a variety of topics, including 
the star formation process, stellar nucleosynthesis and evolution, 
dynamical interaction among stars, and the assembly and evolution of 
galaxies. In particular, open clusters (OCs), which cover a wide range of ages 
and metallicities, have been widely used to constrain the formation and evolution of the 
Milky Way, and more specifically of the Galactic disc 
\citepads[e.g.][]{2010A&A...511A..56P,2013ApJ...777L...1F}. 
This is because some of 
their features, such as ages and distances, can be more accurately determined than 
for field stars.

Open clusters are therefore among the most important 
targets of ongoing and forthcoming Galactic surveys. The \textit{Gaia} mission 
\citepads{2001A&A...369..339P,2005ESASP.576....5M,2005ESASP.576...29L}, for example, will 
provide accurate parallaxes and proper motions for all stars 
down to magnitude 20. Low- and 
medium-resolution spectroscopic surveys, such as the RAdial Velocity Experiment 
\citepads[RAVE;][]{2014A&A...562A..54C} and the Sloan Extension for Galactic 
Understanding and Exploration \citepads[SEGUE;][]{2008AJ....136.2050L}, provide radial velocities, together with some information about the 
chemical content of the stars. High-resolution spectroscopic surveys, 
such as the Apache Point Observatory Galactic Evolution Experiment 
\citepads[APOGEE;][]{2013ApJ...777L...1F}, the Gaia-ESO Survey 
\citepads[GES;][]{2014A&A...561A..94D} and GALactic Archaeology with HERMES 
\citepads[GALAH;][]{2014IAUS..298..322A},  supply accurate radial velocities and detailed 
chemical abundances. Additionally, some projects have been designed 
specifically to only investigate open clusters. For example, the Bologna Open Clusters 
Chemical Evolution project \citepads[BOCCE;][]{2006AJ....131.1544B} uses both 
comparison between observed colour-magnitude diagrams and stellar evolutionary 
models, and the analysis of high-resolution spectra to infer cluster 
properties such as age, distance, and chemical composition. Another survey 
devoted exclusively to OCs is Open cluster Chemical Abundances from Spanish 
Observatories \cite[OCCASO;][]{2014arXiv1412.3509C}. Following a similar strategy to  
GES, this survey obtains accurate radial velocities and chemical abundances 
from high-resolution spectroscopy. 
Taken together, these surveys will produce a breakthrough in our understanding 
of Galactic OCs over the coming decade.

\begin{figure}
\centering
\includegraphics[height=\columnwidth,width=\columnwidth]{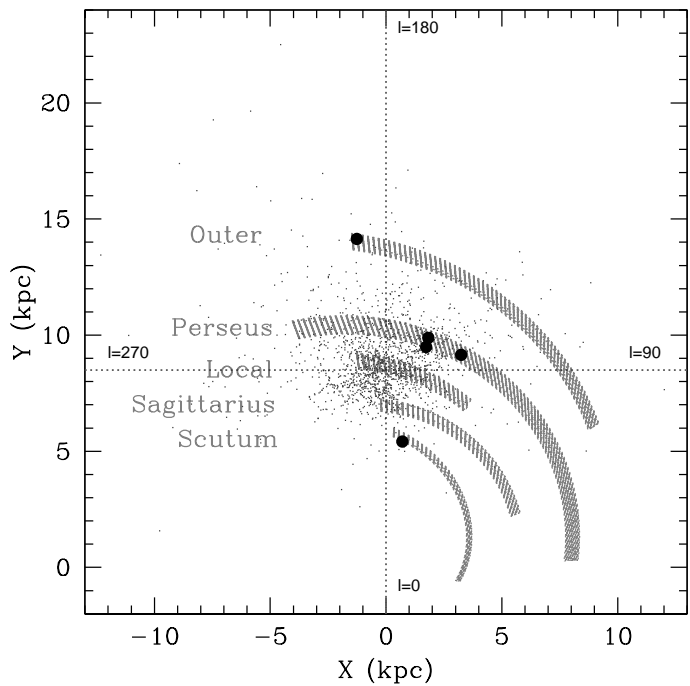}
\caption{Spatial distribution of the five clusters studied in this paper (filled 
circles) superimposed on known OCs (dots) from 
\citetads{2002A&A...389..871D}. The Galaxy's spiral arm positions (dashed areas), obtained from 
\citetads{2014ApJ...783..130R}, have been plotted for reference. The Sun's position is shown by the intersection of the horizontal
and vertical dotted lines. The 
location of the OCs studied here matches the position of some of the Galaxy's 
spiral arms (see Sect. \ref{sec5.2} for discussion).}
\label{fig_dist}%
\end{figure}

However, most of these surveys -- particularly the high-resolution 
spectroscopic ones -- are hampered by a lack of information on cluster 
membership. Only a third of the approximately 2100 OCs known in our Galaxy 
\citepads{2002A&A...389..871D}\footnote{The updated
version of this catalogue can be found at {\href{http://www.astro.iag.usp.br/ocdb/}{\tt
http://www.astro.iag.usp.br/ocdb/}}.}
have radial velocity and/or proper motion information to permit 
membership determination 
\citepads[e.g.][]{2006A&A...446..949D,2007AN....328..889K}. 
\citetads[][hereafter Paper I]{2012A&A...544A.109C} derived radial velocities 
and metallicities from medium-resolution spectroscopy in four previously 
unstudied open clusters. With the same goal, in the current study we have acquired medium-resolution spectra for objects in the lines of sight
of another five poorly studied open clusters. Our sample includes NGC~6603, 
one of the nearest known clusters to the Galactic centre, with a Galactocentric distance of 
5.5 kpc, and Berkeley~23, one of the most distant known OCs located at around 14.2 kpc. The other three clusters studied are 
King~1, NGC~559, and NGC~7245, located near the Perseus spiral arm at about 10 
kpc. Their spatial distribution is shown in 
Fig.~\ref{fig_dist}, and their properties, listed in Table~\ref{clusterspropierties}, are described 
in depth in Sect.~\ref{sec4}.

This paper is organized as follows. Target selection, 
observations, and data reduction are described in Sect.~\ref{sec2};  the membership selection and
procedures used to determine the radial velocities and metallicities are 
explained in Sect.~\ref{sec3}; the results obtained for each cluster are 
presented in Sect.~\ref{sec4}; these results are discussed in Sect.~\ref{sec5} in 
the context of the trends described by OCs in the Galactic disc, and in relation to the spiral arms. Finally, our 
main conclusions are given in Section~\ref{sec6}.

\section{Target selection, observations, and data reduction}\label{sec2}

\begin{figure}
\centering
\includegraphics[height=\columnwidth,width=\columnwidth]{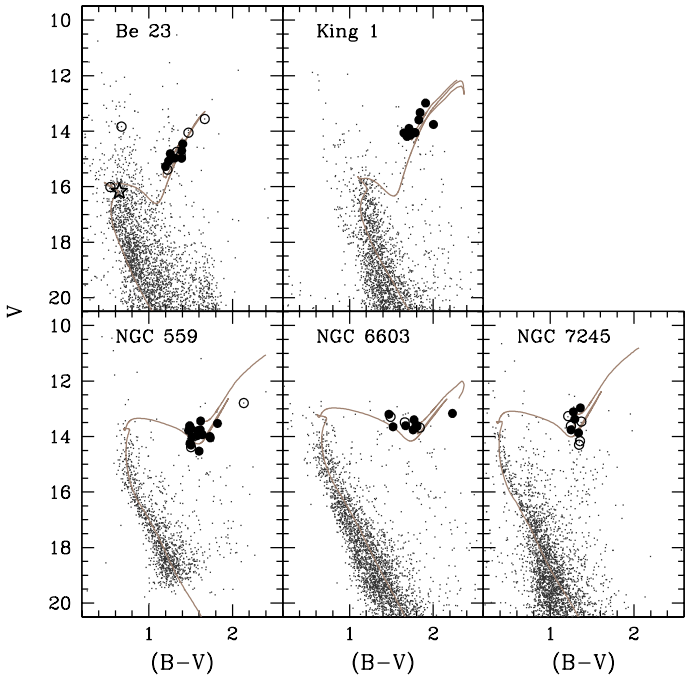}
\caption{Colour--magnitude diagrams of the clusters studied (small grey dots). 
The literature sources for the photometric data are given in the text. Filled circles are giant stars 
in the RGB or RC belonging to the cluster based on their
radial velocities and thus used in the metallicity determination. The open 
star is a main sequence star in Berkeley~23 based on its 
radial velocity, but it has not been used in our metallicity analysis. Open 
circles are
stars observed but classified as non-members based on their radial velocities 
(see 
Sect.~\ref{sec3}). Isochrones derived by \citetads{2004ApJ...612..168P} have 
been plotted as reference using the values listed in 
Table~\ref{clusterspropierties} and the metallicities derived in this paper. 
The RC position is covered by the observed stars.}
\label{CMDs}%
\end{figure}

Our targets were selected amongst the stars located in the expected position 
of the red giant branch (RGB) and the red clump (RC) in the colour--magnitude diagram 
of each cluster.  For this purpose, we used colour--magnitude diagrams from 
\citetads{2011MNRAS.416.1077C}, \citetads{2004BASI...32..371L},
\citetads{2014MNRAS.437..804J}, \citetads{1998MNRAS.299....1S}, and 
\citetads{2007MNRAS.377..829S} for
Berkeley~23, King~1, NGC~559, NGC~6603, and NGC~7245, respectively
(Fig.~\ref{CMDs}). The $BVI$ magnitudes and coordinates of 
the clusters studied were obtained from the WEBDA\footnote{\href{http://webda.physics.muni.cz}{\tt 
http://webda.physics.muni.cz}} database \citepads{1995ASSL..203..127M}, except for NGC~559, for which $BV$ 
photometry and positions were provided directly by Y. C. Joshi (private communication). 
The $K_\mathrm{S}$ magnitudes were extracted from the Two Micron All-Sky Survey
\citepads[2MASS; ][]{2006AJ....131.1163S} database\footnote{Available at \href{http://irsa.ipac.caltech.edu
}{\tt http://irsa.ipac.caltech.edu/}.}.
In total, we observed 64 stars: 18 in 
NGC~559, 15 in
Berkeley~23, 11 in NGC~6603, and 10 in King~1 and NGC~7245.
Table~\ref{obsstars} lists the coordinates, magnitudes, exposure times, and 
total signal-to-noise  ratios per pixel for each target star.

Medium-resolution spectra in the region of the near-infrared \ion{Ca}{II} 
triplet (CaT) at $\sim$8500\,\AA~were obtained using the Intermediate Dispersion 
Spectrograph (IDS) mounted at the Cassegrain
focus of the 2.5 m Isaac Newton Telescope (INT) located at  Roque de los
Muchachos Observatory, Spain. Berkeley~23 was observed in service mode 
on 2013 January 17 and 18, and the other clusters 
were observed on 2014 July 18 and 19. In both cases we used the 
R1200R grism centred on 8500\,\AA~and
the RED+2 CCD, providing a spectral resolution of about 8000. With a few 
exceptions, we obtained two exposures for each target with the star shifted along the slit.  
In some cases, it was possible to observe another star together with the main 
target because it was aligned with the slit. The spectra of these additional stars 
were extracted and analysed in the 
same way as those of the main targets.

The data reduction is explained in detail in Paper~I and was performed 
using 
IRAF\footnote{The Image Reduction and Analysis Facility, IRAF, is
distributed by the National Optical Astronomy Observatories, which are operated
by the Association of Universities for Research in Astronomy, Inc., under
cooperative agreement with the National Science Foundation.} packages. 
Briefly, each image is overscan-subtracted, trimmed, and flatfield-corrected with the
\texttt{ccdproc} tool. We did not perform bias subtraction because bias 
level was not constant throughout the night. Since we acquired
two exposures for each target with shifts along the slit, we subtracted 
one from the other, obtaining a positive and a negative spectrum in the same image.
In this way, we are subtracting the sky from the same physical pixel in which
the star was observed, which minimizes the effect of pixel-to-pixel sensitivity
variations. Of course, a time dependency remained because the two spectra had not
been taken simultaneously. These sky residuals were eliminated in the following
step, in which the spectrum was extracted in the traditional way using \texttt{apall} tool, and the
remaining sky background was subtracted from the information on both sides of the
stellar spectra. After this, each spectrum was wavelength-calibrated before the two spectra
(one positive, one negative) were subtracted again to obtain the final spectrum. 
Finally, the spectrum was normalized by fitting a polynomial, 
excluding the strongest lines in the
wavelength range, such as those of the calcium triplet.

\begin{table}
\begin{minipage}[htb]{\columnwidth}
\caption{Adopted cluster parameters.}
\label{clusterspropierties}
\centering
\renewcommand{\footnoterule}{}  
\begin{tabular}{l c c c c c}
\hline\hline       
Cluster   & $E(B-V)$ & $(m-M)_{\rm o}$ & Age & $R_{\rm GC}$ & $z$ \\
         & (mag)   & (mag)     & (Gyr)  & (kpc) & (kpc) \\
\hline
Be~23\footnote{Average of the different values obtained by \citetads{2011MNRAS.416.1077C}. See text for details.}
         & 0.33$\pm$0.08 & 13.8$\pm$0.1 & 1.1$\pm$0.2 & 14.2 & 0.55\\
King~1\footnote{Average of the values by \citetads{2008PASJ...60.1267H} and \citetads{2004BASI...32..371L}}
         & 0.66$\pm$0.06 & 11.5$\pm$0.1 & 2.2$\pm$0.9 & 9.6 & 0.06 \\    
NGC~559\footnote{Average of the values derived by \citetads{2014MNRAS.437..804J}, \citetads{2007A&A...467.1065M} and \citetads{2002JKAS...35...29A}}
         & 0.76$\pm$0.06 & 11.8$\pm$0.1 & 0.4$\pm$0.2 & 10.0 & 0.03 \\
NGC~6603\footnote{Average of the values calculated by \citetads{1998MNRAS.299....1S} and \citetads{1993A&A...270..117B}}
         & 0.67$\pm$0.16  & 12.5$\pm$0.4 & 0.3$\pm$0.2 & 5.5 & $-0.07$ \\
NGC~7245\footnote{Average of the values obtained by \citetads{1997A&A...328..158V}, \citetads{2007AN....328..889K}, 
\citetads{2010AstL...36...14G}, and \citetads{2011AJ....141...92J}}
         & 0.42$\pm$0.02  & 12.6$\pm$0.3 & 0.4$\pm$0.1 & 9.71 & $-0.11$ \\
\hline
\end{tabular}
\end{minipage}
\end{table}
\setcounter{table}{2}

\section{Membership selection, radial velocity and metallicity determination}\label{sec3}

The radial velocity of each star was calculated using the \texttt{fxcor}
task in IRAF, which performs a cross-correlation between the target and template
spectra of known radial velocity. For Berkeley~23, we used as templates the same 
stars used in Paper~I. For the July 2014 run, we used the stars 
NGC~6819~W0968 
and NGC~6705~S5688 as templates observed on the same nights. These stars were
selected because they have similar abundances and spectral types to the 
target stars. Moreover, their radial velocities have been accurately 
determined from high-resolution spectroscopy, and they have not been reported as 
spectroscopic binaries or radial velocity variables 
\citepads{1993AstL...19..232G,2014A&A...569A..17C}. Since the 
radial velocities obtained using the different templates are similar within the 
uncertainties, the final radial velocity for each target star was
obtained as the average of the velocities obtained from each template, weighted
by the width of the correlation peaks. Radial velocities for each 
star are listed in column~9 of Table~\ref{obsstars}. 

\begin{figure}
\centering
\includegraphics[height=\columnwidth,width=\columnwidth]{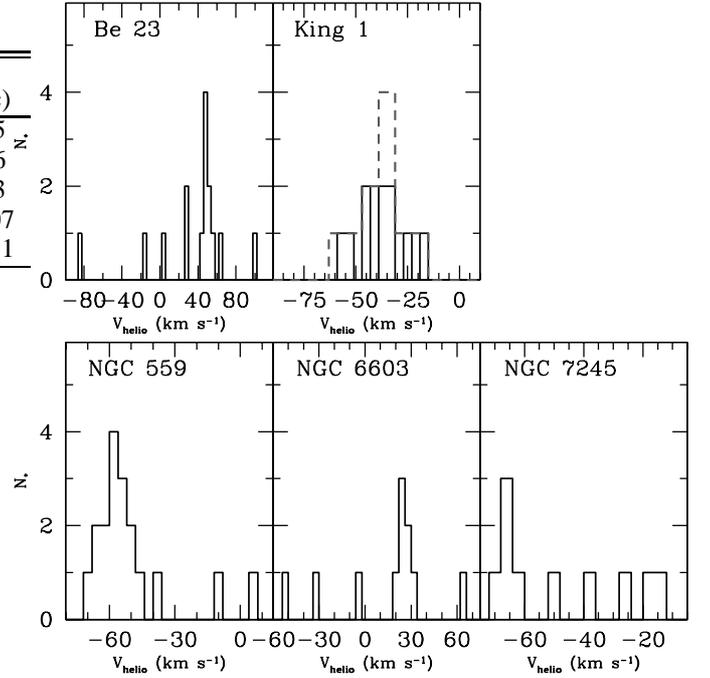}
\caption{Radial velocity distributions of the five OCs studied obtained using a 
bin of 4 km s$^{-1}$. In the case of King~1, the dashed line represents the 
histogram obtained using a bin of 8 km s$^{-1}$. The x-axes are not on the same scale.}
\label{Vrdist}%
\end{figure}

The velocity distribution of 
the stars observed in the area of each cluster is shown in Fig.~\ref{Vrdist} 
and discussed in Section~\ref{sec4}. 
Cluster membership was determined from radial velocities.
An iterative k-sigma clipping was used in the following way. The mean and standard deviation 
of the radial velocities of all the stars in the cluster area were computed. 
Stars with radial velocities outside $\pm$3$\sigma$ were rejected. This procedure was
repeated until no more stars were rejected. The final number of cluster members is listed in the 
last column of Table~\ref{clustersresults}. The average radial velocity 
of each cluster, obtained as the mean of the radial velocities of all cluster members, and its 
standard deviations are listed in column~6 of Table~\ref{clustersresults}.

Metallicities for all RC and RGB stars selected as members from their 
radial velocities were determined from the strengths of their 
CaT lines. As in Paper~I, the equivalent widths of these lines were calculated 
using the procedure described in depth by \citetads{2007AJ....134.1298C}. In 
brief, the equivalent width of each line is determined as the area between the 
line profile and the continuum. The line profile is determined by fitting a Gaussian plus 
a Lorentzian, which 
provides the best fit to the line core and wings. Although the spectrum was 
previously 
normalized, the continuum level is recalculated by performing a linear fit
to the mean values of several continuum 
bandpasses defined for this purpose. The bandpasses used to fit the
line profile and to determine the continuum position are those described by
\citetads{2001MNRAS.326..959C}. The equivalent widths of 
each CaT line and their uncertainties, which were determined for each star, are also listed in 
Table~\ref{obsstars}. Finally, the CaT index, denoted $\Sigma$Ca, was obtained as 
the sum of the equivalent widths of the three CaT lines. 

\begin{figure}
\centering
\includegraphics[height=\columnwidth,width=\columnwidth]{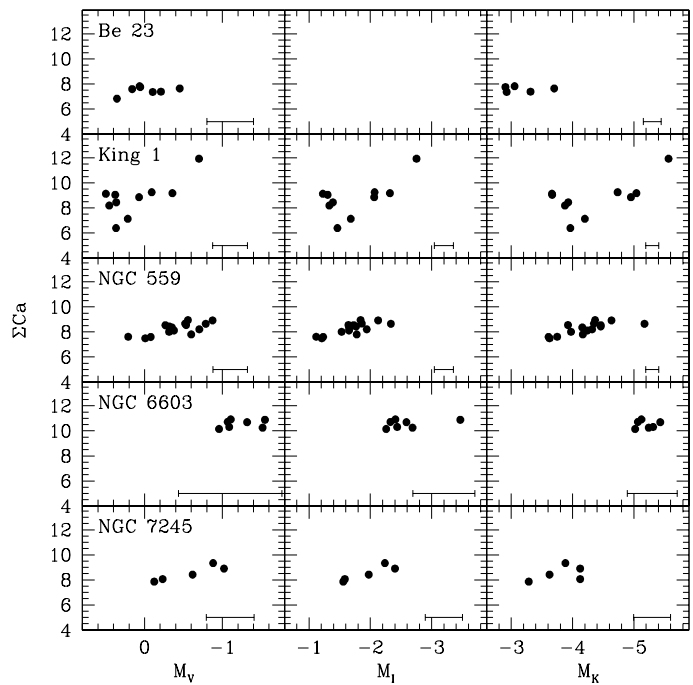}
\caption{Stars selected as RC or RGB members of each cluster in the $M_V$--$\Sigma$Ca (left), $M_I$--$\Sigma$Ca 
(centre), and $M_K$--$\Sigma$Ca (right) planes. The mean sigma for each luminosity indicator is 
shown in bottom right corner of each panel and is mainly due to the 
distance modulus uncertainty. Error bars in $\Sigma$Ca are smaller than the point size. }
\label{fig_mag_sca}%
\end{figure}

The strength of the CaT lines depends not only on the chemical abundance 
but also on the temperature and gravity of the star. This dependence is removed 
because stars of the same chemical composition define a clear 
sequence in the luminosity--$\Sigma$Ca plane when the temperature and/or gravity 
change. Several luminosity indicators have been used in the literature for this 
purpose. One of the most widely used is the magnitude of a star relative to the 
position of the horizontal branch (HB) in the $V$ filter, denoted  $V-V_{\rm HB}$,
which also removes any dependence on distance and reddening. An alternative approach is 
to use the absolute magnitude in different bandpasses, such as $V$, $I$, and $K_\mathrm{S}$, 
denoted  $M_V$, $M_I$, and $M_K$, respectively. Owing to the difficulty 
of defining the HB position in poorly populated clusters, we used the 
second approach. The absolute magnitudes of each star were 
obtained from the distance modulus and reddening listed in Table~\ref{clusterspropierties} using the 
extinction coefficients listed in Table 6 from \citetads{1998ApJ...500..525S}. The position of the 
member stars in each cluster in the $M_V$--$\Sigma$Ca (left), $M_I$--$\Sigma$Ca 
(centre), and $M_K$--$\Sigma$Ca (right) planes is shown in Figure~\ref{fig_mag_sca}.

In Paper~I, metallicities were calculated using the relationships derived by 
\citetads{2007AJ....134.1298C}, assuming a linear relation between 
the CaT index and the luminosity indicators. However, 
\citetads{2010A&A...513A..34S} have demonstrated that this assumption is not valid, 
particularly for metal-poor regimes. For this reason, 
\citetads{2013MNRAS.434.1681C} recomputed the 
\citetads{2007AJ....134.1298C} relationships, introducing two additional terms to 
account for the non-linearity effects.
In any case, these authors have demonstrated that both calibrations 
produce similar results, within $\pm 0.2$ dex, in the range of metallicities 
expected for open clusters. We used the most 
recent relationships derived by \citetads{2013MNRAS.434.1681C} to obtain the 
metallicities of the observed clusters. The individual values obtained for each 
star and bandpass are listed in Table~\ref{obsstars}. The metallicity in each bandpass
is the average of all the stars in the cluster, and the associated uncertainty is the
standard deviation. Finally, the metallicity of each 
cluster was obtained as the mean of the values obtained from each bandpass. 
 Also, the associated uncertainties were computed as the mean of the 
uncertainties derived from each bandpass. This value is a better estimate of
the uncertainty than the standard deviation of the metallicities 
calculated from each bandpass since they are not independent determinations.
The mean values are listed in 
Table~\ref{clustersresults} and discussed below.

In this paper we have 
derived metallicities in both RC and RGB stars using relationships suitable for
RGB stars. \citetads{2012AJ....143...44D} have studied the behaviour of the 
strength of the CaT lines as a function of temperature and gravity from synthetic 
spectra. Their Fig.~9 predicts a slight decrease in $\Sigma$Ca with temperature 
in the metallicity range covered by open clusters. 
Since RC stars are slightly hotter than RGB ones, an RC star with the same 
magnitude as a RGB one may have a slightly higher $\Sigma$Ca. Taking
NGC~559 as an example and using the 
relationships derived by \citetads{2012AJ....143...44D}, the maximum difference in 
$\Sigma$Ca between RC and RGB stars is $\sim$0.16 \AA. 
This implies that the metallicity of an RC star derived with the RGB 
relationships may be overestimated by a maximum of $\sim$0.07 dex, independent of the 
luminosity indicator used. Although systematic, this value is lower than
the uncertainties of the relationships used and lower than the 
standard deviation of the metallicity determination of each cluster.

\begin{table*}
\begin{minipage}[t]{17.5cm}
\caption{Mean cluster metallicities and radial velocities and their standard deviation.
The last column lists the number of member stars.}
\label{clustersresults}
\centering
\renewcommand{\footnoterule}{}  
\begin{tabular}{l c c c c c c }
\hline\hline       
Cluster   &  [Fe/H]$_V$ & [Fe/H]$_I$ & [Fe/H]$_K$ & $\langle$ [Fe/H] $\rangle$\footnote{Mean of the values obtained in each bandpass. The adopted uncertainty is the mean of the uncertainties of the metallicity determination in each bandpass.} & $\langle V_{\rm r} \rangle$\footnote{Mean and standard deviation of the radial velocities of the stars selected as cluster members.} & \# \\
         &         &       &       &       & (km s$^{-1}$) & \\
\hline
Be~23 & $-$0.42$\pm$0.13 & & $-$0.43$\pm$0.13 & $-$0.42$\pm$0.13 & 48.6$\pm$3.4 & 8\footnote{Seven red giant and one main sequence stars, the latest not used in the metallicity determination.} \\
King~1 & +0.13$\pm$0.51 & +0.00$\pm$0.52 & $-$0.16$\pm$0.53 & $-$0.01$\pm$0.52 & $-$38.4$\pm$11.6 & 10\footnote{The number of stars used in our analysis. However, we cannot ensure that we have sampled real cluster members for the reasons described in Sect. 4.2} \\       
NGC~559 & $-$0.22$\pm$0.14 & $-$0.18$\pm$0.14 & $-$0.35$\pm$0.14 & $-$0.25$\pm$0.14 & $-$58.4$\pm$6.8 & 15 \\
NGC~6603 & +0.42$\pm$0.13 & +0.53$\pm$0.14 & +0.33$\pm$0.17 & +0.43$\pm$0.15 & 26.0$\pm$4.3 & 7 \\
NGC~7245 & $-$0.17$\pm$0.14 & $-$0.14$\pm$0.17 & $-$0.15$\pm$0.24 & $-$0.15$\pm$0.18 & $-$65.3$\pm$3.2 & 5 \\
\hline
\end{tabular}
\end{minipage}
\end{table*}

\section{Cluster-by-cluster discussion}\label{sec4}

\subsection{Berkeley~23}

Berkeley~23 is an OC located in the third Galactic quadrant towards  
the Galactic anticentre. Its colour--magnitude diagram is strongly contaminated by field stars
and its sequence is poorly defined. However, a broad and clumpy main sequence is 
still observed. Moreover, there is a mild excess of stars in the region where the RC is expected (e.g. $B-V\sim1.2$ and $V\sim15.0$).
There are only three studies of this distant cluster in 
the literature. By isochrone fitting of \textit{UBVI} colour--magnitude diagrams,
\citetads{2002AJ....123..905A} found an age of 790$\pm$160 Myr, a distance 
modulus of $(m-M)_{\rm o}$=~14.2$\pm$0.2 mag, a reddening of $E(B-V)$=~0.40$\pm$0.05 mag, 
and [Fe/H] =~0.07. Also, using isochrone fitting in \textit{BVI} colour--magnitude diagrams,
\citetads{2004PASJ...56..295H} derived an age of 1.8~Gyr, a distance 
modulus $(m-M)_{\rm o}$=~13.81~mag, a reddening  $E(B-V)$=~0.30, and 
 metallicity  [Fe/H]$\sim-0.7$. More recently, 
\citetads{2011MNRAS.416.1077C} have obtained
a \textit{BV} colour--magnitude diagram for Berkeley~23 as part of the BOCCE 
project. By comparing with stellar evolutionary models, they 
determined an age between 1.1 and 1.3~Gyr, a distance modulus $(m-M)_{\rm o}$ between 13.6 and 14.0~mag, a 
reddening $E(B-V)$ between 0.225 and 0.425~mag, and a metallicity between solar and 
half solar, i.e. [Fe/H] $\sim-0.3$. The uncertainties in the parameters were
caused by the difficulties in adjusting the main sequence turn-off and the RC at the same time.
In general, the quantities derived by the 
three studies agree within the large uncertainties except for the metallicity. 
Here, we use the mean of the values 
derived by \citetads{2011MNRAS.416.1077C} as distance modulus and reddening. They lie between the values derived 
by \citetads{2002AJ....123..905A} and \citetads{2004PASJ...56..295H}.

A total of fifteen stars have been observed in the Berkeley~23 area. Only 
eight of them have similar radial velocities defining a clear peak in the 
radial velocity distribution (top left panel of Fig.~\ref{Vrdist}). From these 
eight stars we derived a mean radial velocity of 
$\left\langle V_{\rm r}\right\rangle = 48.6\pm3.4$ km s$^{-1}$.

One of the eight stars selected as members is a main sequence 
star (Fig.~\ref{CMDs}) and was not used in the metallicity determination. The 
average metallicity of the seven giant stars is [Fe/H]$ =-0.42\pm0.13$  
and [Fe/H] $=-0.43\pm$0.13  using $V$ and $K_\mathrm{S}$ magnitudes, respectively. There 
are no $I$ magnitudes  in the literature. Our result is in very good agreement 
with the value ([Fe/H]$\sim-0.3$) obtained from comparison between 
observational and synthetic colour--magnitude diagrams
by \citetads{2011MNRAS.416.1077C}. Our metallicity 
determination also agrees with the value ([Fe/H] $\sim-0.7$)
obtained by \citetads{2004PASJ...56..295H} from isochrone 
fitting, when taking the associate uncertainty of about $\pm0.3$ in their method into account 
 (corresponding to the separation in metallicity of the isochrones 
used).

\subsection{King~1}

King~1 is an old OC located in the second Galactic quadrant in the direction of the 
Galactic anticentre.  
To our knowledge, three studies in recent years have derived its physical 
properties from 
colour--magnitude diagrams. The King~1 colour--magnitude diagram shows a broad 
main sequence with a well-populated RC. \citetads{2004BASI...32..371L} used 
\textit{UBVRI} 
photometry and isochrone fitting to obtain an age of $1.6\pm0.4$ Gyr, a distance 
modulus of $(m-M)_{\rm o}=11.38$ mag, and a reddening of $E(B-V)=0.70\pm0.05$, when assuming 
solar metallicity. Using the same technique and \textit{BV} wide-field 
photometry, \citetads{2007A&A...467.1065M} obtained an age of $\sim$4~Gyr, a 
distance modulus of $(m-M)_{\rm o}=10.17^{+0.32}_{-0.51}$ mag, and a reddening of 
$E(B-V)=0.76\pm0.09$ mag, also assuming solar metallicity. Finally, from \textit{VI} photometry 
\citetads{2008PASJ...60.1267H} derived an age of 
2.8 Gyr with a distance modulus $(m-M)_{\rm o}=11.57$ and reddening $E(B-V)=0.62$ mag, 
when assuming solar metallicity. The values obtained by 
\citetads{2004BASI...32..371L} and \citetads{2008PASJ...60.1267H} agree within 
the uncertainties. However, the distance modulus and age derived by 
\citetads{2007A&A...467.1065M} are very different from those obtained by the 
other two studies. For this reason,  in our work we use the distance modulus 
and reddening obtained by averaging the values derived by 
\citetads{2004BASI...32..371L} and \citetads{2008PASJ...60.1267H}, as listed in 
Table~\ref{clusterspropierties}, although we have investigated the impact of 
using the values derived by \citetads{2007A&A...467.1065M} in our results.

The ten stars observed in the King~1 area do not outline a well-defined 
distribution with the velocity bin of 4 km s$^{-1}$ used to construct the 
velocity histograms of the other clusters (solid black histogram in 
middle top panel of Figure~\ref{Vrdist}). Therefore, we have doubled the 
size to 8 km s$^{-1}$, but even so we find that a clear peak is not 
obtained. In this case none of the observed stars can be rejected from its 
radial velocity with the k-sigma clipping procedure used in the other 
clusters. The average velocity of the ten stars observed in the King~1 area is $\langle V_{\rm r}
\rangle=-38.4$ km s$^{-1}$, with a dispersion of $\pm11.6$  km s$^{-1}$.

The other clusters we studied have relatively well-defined sequences in the
luminosity--$\Sigma$Ca planes (Fig.~\ref{fig_mag_sca}). However, the ten stars observed in King~1 are more sparsely 
distributed. From these stars we obtained [Fe/H] $=+0.13\pm0.51$, 
[Fe/H] $=+0.00\pm0.52$ and [Fe/H] $=-0.16\pm0.53$, 
using $V$, $I$, and $K_\mathrm{S}$, respectively. If the distance modulus and reddening 
derived by \citetads{2007A&A...467.1065M} are used, the metallicities obtained 
are higher: [Fe/H] $=+0.44\pm0.54$, 
[Fe/H] $=+0.29\pm0.56$, and [Fe/H] $=+0.03\pm0.56$, 
using $V$, $I$, and $K_\mathrm{S}$, respectively. These values seem to be too high for a cluster at 
the Galactocentric distance of King~1. In both cases, the derived uncertainties are 
twice the values obtained for the other clusters because of the high dispersion 
in the luminosity--$\Sigma$Ca planes, as seen in Figure~\ref{fig_mag_sca}. 
This and the lack of a clear peak in the velocity distribution, 
may imply that the number of real cluster members in our sample is very small, or even zero.
Therefore, both the mean radial velocity and metallicity values should be
treated with caution.

\subsection{NGC~559}

NGC~559 is a moderately populated OC of the second Galactic quadrant located in the 
vicinity of the Perseus arm (Fig.~\ref{fig_dist}). Its colour--magnitude diagram 
shows a well-populated main sequence with a diffuse but still clear RC. 
Several studies have been devoted to determining its properties, mainly through the use of 
isochrone fitting 
\citepads[e.g.][]{1969ArA.....5..221L,1975MNRAS.172..681J,1975A&AS...21...99G, 
2002JKAS...35...29A,2007A&A...467.1065M,2014MNRAS.437..804J}. The reddening 
towards NGC~559 is high with a mean value of $E(B-V)=0.76\pm0.06$ mag, obtained 
after averaging determinations from recent studies based on CCD photometry 
\citepads{2002JKAS...35...29A,2007A&A...467.1065M,2014MNRAS.437..804J}. In the 
same way, the mean distance modulus of NGC~559 is $(m-M)_{\rm o}=11.80\pm0.12$ mag. 
The age values determined by these studies are $225\pm25$ Myr, obtained 
by \citetads{2014MNRAS.437..804J}, and 630 Myr, derived by 
\citetads{2007A&A...467.1065M}, with a mean value of $400\pm200$ Myr. All these 
values were obtained assuming solar metallicity.

Only three of the 18 stars observed do not seem to be members of 
NGC~559 from their radial velocity. The remaining 15 stars define a clear 
radial velocity distribution (Fig.~\ref{Vrdist}) with a mean value of 
$\langle V_{\rm r} \rangle=-58.4\pm6.8$  km s$^{-1}$.

From these 15 stars we have derived [Fe/H] $=-0.22\pm0.14$, [Fe/H] $=-0.18\pm0.14$, and 
[Fe/H] $=-0.35\pm0.14$ in the $V$, $I$, and $K_\mathrm{S}$ bandpasses, respectively. These 
results are in good agreement with the metallicity estimate obtained by 
\citetads{2002JKAS...35...29A} from isochrone fitting ([Fe/H] $\sim-0.32$).
Other studies \citepads[e.g.][]{2007A&A...467.1065M,2014MNRAS.437..804J} 
have assumed solar metallicity.

\subsection{NGC~6603}

NGC~6603 is a rich OC located in the inner disc towards the Galactic centre in 
Sagittarius. Its colour--magnitude diagram shows  features that are typical of an 
intermediate-age OC with the clear presence of RC stars. A few studies have 
been devoted to determining its physical parameters using photometry 
\citepads[e.g.][]{1993A&A...270..117B,1998MNRAS.299....1S,
2005A&A...438.1163K} and integrated spectra 
\citepads[e.g.][]{1993MNRAS.260..915S,1993A&A...270..117B}. These studies agree with a line-of-sight reddening between $E(B-V)=0.50$ \citepads{2005A&A...438.1163K} 
and $E(B-V)=0.56$ \citepads{1998MNRAS.299....1S} in the direction of
 NGC~6603  and an internal reddening of 
$E(B-V)=0.29$ mag \citepads{1993A&A...270..117B}. The distance modulus is 
constrained between $(m-M)_{\rm o}=12.16$ \citepads{1998MNRAS.299....1S} and 
$(m-M)_{\rm o}=12.78$ \citepads{1993A&A...270..117B} with a mean value of 
$(m-M)_{\rm o}=12.41\pm0.32$ mag. The age determined for NGC~6603 is between 
$200\pm100$ Myr \citepads{1993A&A...270..117B} and 500 Myr 
\citepads{1998MNRAS.299....1S}, although \citetads{2005A&A...438.1163K} assigned 
an age of 60 Myr to this cluster. To our knowledge, NGC~6603 is the only cluster 
in our sample in which spectra of individual stars have been previously acquired. 
\citetads{2008AJ....136..118F} acquired medium-resolution ($R\sim15\,000$) spectra 
for 55 stars in NGC~6603. From proper motions and radial velocities, 
they concluded that only four of them were NGC~6603 members with a mean 
radial velocity of $\langle V_{\rm r} \rangle=21.34\pm0.92$ km s$^{-1}$.

Of the 11 stars observed in our sample, only seven are cluster members, 
the other four having discordant radial velocities. 
We derived a mean radial velocity of $\langle V_{\rm r} \rangle=26.0\pm4.3$ km s$^{-1}$. 
This value agrees, within the uncertainties, with the mean radial velocity found
by \citetads{2008AJ....136..118F}. A direct comparison with their results is not possible because 
there are no stars in common.

For NGC~6603 we derived a mean metallicity of 
[Fe/H] $=+0.42\pm0.13$, $+0.53\pm0.14$, and $+0.33\pm0.17$ from $V$, $I$, and 
$K_\mathrm{S}$ magnitudes, respectively. This places NGC~6603 within the most metal-rich 
OCs known. Of course, this result has to be confirmed by 
high-resolution spectroscopy. All previous studies 
of this cluster have assumed a solar metallicity 
\citepads[e.g.][]{1998MNRAS.299....1S,2005A&A...438.1163K}, which does not seem to be the case.

\subsection{NGC~7245}

NGC~7245 is a sparse cluster located in the second Galactic quadrant. Its 
colour--magnitude diagram shows a well-defined main sequence with a sparse concentration 
of stars in the expected position of the RC (e.g.\ $B-V\sim1.3$ and $V\sim13.6$). One of the first 
photometric studies in this cluster 
\citetads {1970A&A.....8..213Y} using the \textit{RGU} system found a distance modulus  
$(m-M)_{\rm o}=11.42$ and reddening $E(B-V)\sim0.60$ mag. More 
recently, and on the basis of \textit{BV} Johnson CCD photometry and isochrone fitting,
\citetads{1997A&A...328..158V} derived an age of 320 Myr, distance 
modulus $(m-M)_{\rm o}=12.23\pm0.15$, and reddening $E(B-V)=0.40\pm0.02$ mag, 
assuming solar metallicity. From \textit{UBV} CCD photometry 
\citetads{2007AN....328..889K} find an age of 400 Myr, distance modulus 
$(m-M)_{\rm o}=12.9\pm0.2$ mag, and reddening $E(B-V)=0.45\pm0.02$ mag, assuming 
solar metallicity. \citetads{2010AstL...36...14G} used \textit{BVR$_c$I$_c$} 
photometry to estimate an age of $320\pm40$ Myr with a distance modulus
$(m-M)_{\rm o}=12.65\pm0.13$ and reddening $E(B-V)=0.42\pm0.03$ mag, 
assuming solar metallicity. Finally, 
\citetads{2011AJ....141...92J} derived an age of 445 Myr with a distance 
modulus $(m-M)_{\rm o}=12.7$ and reddening $E(B-V)=0.45$ mag from \textit{VI} photometry, also 
assuming solar metallicity. Since all studies based on 
CCD photometry produce relatively similar results within the uncertainties, we 
used as reddening and distance modulus the average of the values 
obtained by \citetads{1997A&A...328..158V}, \citetads{2007AN....328..889K}, 
\citetads{2010AstL...36...14G}, and \citetads{2011AJ....141...92J}, listed in 
Table~\ref{clusterspropierties}.

Five of the ten stars observed in NGC~7245 define a clear peak in the velocity 
distribution (bottom right panel of Fig.~\ref{Vrdist}). The remaining stars have 
discordant radial velocities. The 
mean radial velocity of the five stars selected as members is $\langle 
V_{\rm r} \rangle=-65.3\pm3.2$  km s$^{-1}$.

From the NGC~7245 members, we obtained an average metallicity of 
[Fe/H] $=-0.17\pm0.14$, [Fe/H] $=-0.14\pm0.17$, and [Fe/H] $=-0.15\pm0.24$ 
using $V$, $I$, and $K_\mathrm{S}$, respectively. There 
are no metallicity determinations of this cluster in the literature, and all 
studies using isochrone fitting have assumed a solar content.

\section{Discussion}\label{sec5}

\begin{figure}
\centering
\includegraphics[height=10cm,width=\columnwidth]{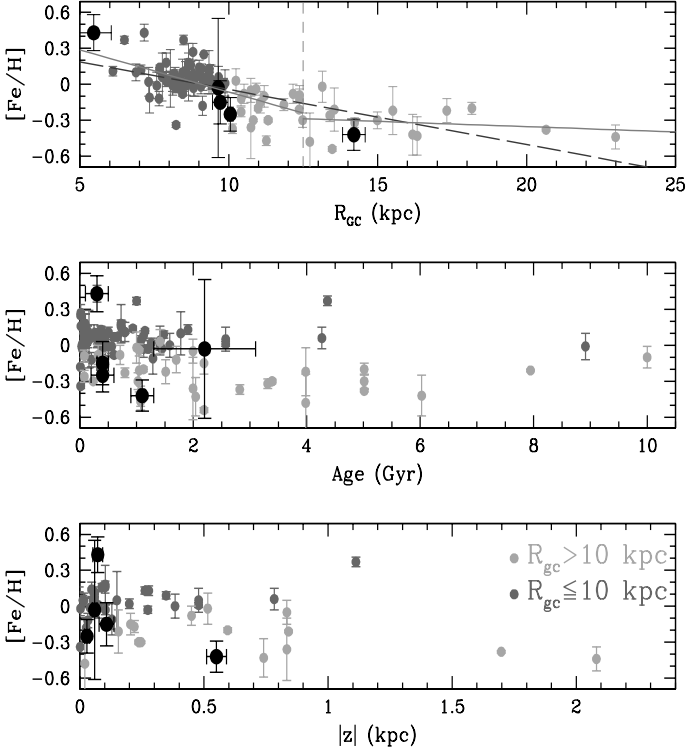}
\caption{Gradient in [Fe/H] as a function of $R_{\rm GC}$ (top), age (middle), and $|z|$ (bottom) of
the OCs in the \citetads{2011A&A...535A..30C} compilation. For the clusters 
studied here, we have used the weighted mean of the values obtained in the three 
bands listed in column~5 of Table~\ref{clustersresults}. Dark and light grey
points are clusters inside and outside 
radial distances 
of 10 kpc, respectively. Black
points are the clusters studied in this paper. Solid lines in the top panel 
represent two separate linear fits obtained by 
\citetads{2011A&A...535A..30C} for OC inside and outside 12.5 kpc. The dashed line is the fit obtained 
by the same authors using all open clusters.}
\label{trends}%
\end{figure}

\subsection{Comparison with trends in the Galactic disc}\label{sec5.1}

Observed trends described by OCs in the Galactic disc (e.g.\ 
metallicity with Galactocentric distance, $R_{\rm GC}$, vertical distance to the 
Galactic plane, $|z|$, and age) indicate that the chemical composition of OCs 
is determined mainly by their location in the disc and not by the moment when they were formed \citepads[e.g.][]{2011A&A...535A..30C}. It is therefore 
helpful to compare the clusters studied here with the trends described by the bulk 
of OCs in order to check that they share features. 
This comparison is even more important in our case because our sample 
includes clusters in  poorly sampled regions of the disc, such as the innermost 
and outermost areas. To do this comparison we used the compilation of
clusters obtained by \citetads{2011A&A...535A..30C}. Briefly, their sample 
includes chemical abundances available in the literature derived from 
high-resolution spectroscopy ($R\geq$18\,000). We refer the reader to that
paper for more details. The run of metallicity with $R_{\rm GC}$, age, and $|z|$ for 
this sample has been plotted in the top, middle, and bottom panels of Fig.~\ref{trends},
respectively. The top panel shows linear fits to the sample,
when assuming a change in the slope at $R_{\rm GC}\sim$12.5 kpc, the linear fit to the whole sample. It also plots clusters in the inner 10 kpc and
from there outwards in the three panels.

The five clusters studied here closely follow  the trends observed in the Galactic disc. 
Berkeley~23 has a similar metallicity to other OCs located at the same 
Galactocentric distance. NGC~6603 is located even farther in than the innermost OCs 
in the \citetads{2011A&A...535A..30C} sample. If the high metallicity obtained 
here for this cluster is confirmed by high-resolution data, this may
imply that the metallicity gradient in the inner disc is even steeper than previously 
thought.
The three other OCs match the observed trends even when considering the 
large uncertainties in the metallicity determination of King~1. NGC~559, NGC~7245, and King~1 are located almost in the 
Galactic plane, $|z|\leq$ 120 pc, with metallicities similar to other coeval OCs 
located at the same vertical distances. Only Berkeley~23 has a considerable 
vertical distance from the Galactic plane, in agreement with other clusters 
located at similar Galactocentric distances. All of them match the 
age--metallicity distribution described by most  open clusters. 

Finally, the high 
metallicity of NGC~6603 seems to place this cluster outside the age and 
distance from the plane distributions defined by most of the open clusters. There 
are at least three known OCs with [Fe/H] $\geq0.3$. These are 
NGC~6253, NGC~6583 and NGC~6791. Two of them (NGC~6253 and NGC~6583) are located 
at similar distances from the Galactic plane, although they are situated at slightly different 
Galactocentric distances from NGC~6603. 
Moreover, NGC~6253 has a similar age to NGC~6603, whereas
NGC~6583 is slightly older. Only NGC~6791 differs from the other three very metal-rich OCs: 
it is older, more distant,
and  is located more than 1~kpc above the Galactic plane.

Summarizing, the values derived 
for the five clusters studied here match the metallicities of other OCs with similar 
ages and locations in the Galactic plane.

\subsection{Relation to spiral arms}\label{sec5.2}

From Fig.~\ref{fig_dist} it is noteworthy that King~1, NGC~559, and 
NGC~7245 are located in the Perseus spiral arm and Berkeley~23 in 
the outer spiral arm. NGC~6603 is located near the Scutum arm. 
It is therefore interesting to investigate the relation of these clusters with 
such arms.

The first three columns of Table~\ref{table:vrad} list the heliocentric radial velocities induced
by differential Galactic rotation and the peculiar motion of the 
Sun with respect to the local standard of rest (LSR) projected onto
the line of sight of each cluster, $V_{\rm r}^{\rm GC}$ using the Sun's location and 
Galactic rotation computed or adopted by \cite{2014ApJ...783..130R}, \cite{2011MNRAS.418.1423A}, and
\cite{2009PASJ...61..227S}. The peculiar 
motion of the Sun has been adopted as $U_\odot=10.7$,
$V_\odot=15.6$, and $W_\odot=7.0$ km s$^{-1}$ \citep{2014ApJ...783..130R}.
 The error margins of $V_{\rm r}^{\rm GC}$ were computed
using the $\pm\sigma_{(m-M)_{\rm o}}$ values listed in Table~\ref{clusterspropierties}.
For NGC~6603, the distance modulus uncertainty
introduces more variation than the difference among adopted Galactic parameters.
For King~1, NGC~559, and NGC~7245, the three clusters in the Perseus arm, all 
results are marginally compatible. Finally, for Berkeley~23, the difference
in Galactic parameters is larger than the difference due to the distance 
modulus uncertainty. Adopting the same values as \cite{2011MNRAS.418.1423A},
which results in intermediate values for $V_{\rm r}^{\rm GC}$, the peculiar
radial velocities of the clusters with respect to their LSR were computed
as $V_{\rm r}^{\rm pec}=\langle V_{\rm r}\rangle -V_{\rm r}^{\rm GC}$. The values are also
listed in Table~\ref{table:vrad} (last column). The errors are computed as 
$\sqrt{\sigma_{\langle V_{\rm r}\rangle}^2+\sigma_{V_{\rm r}^{\rm GC}}^2}$.
Of the three clusters
located in the Perseus arm, NGC~559 and NGC~7245 have the same age and 
$V_{\rm r}^{\rm pec}$, and they have compatible [Fe/H] values. Taken together, these values
may indicate a relationship in the formation
of these open clusters. 
The peculiar radial 
velocities also roughly agree with the values found by
\cite{2014ApJ...783..130R} for a set of high-mass star-forming regions
in the same arms.

\begin{table}[h]
\caption{Heliocentric radial velocities computed using three different Sun's 
locations and Galactic rotations and the peculiar radial with respect to the 
local standard of rest (LSR). Units are km s$^{-1}$.
}
\label{table:vrad}
\begin{tabular}{lcccc}
\hline
Cluster       & $V_{\rm r}^{\rm GC(1)}$ &$V_{\rm r}^{\rm GC(2)}$ & $V_{\rm r}^{\rm GC(3)}$ &$V_{\rm r}^{\rm pec}$ \\
\hline
{\em Scutum:} \\
NGC~6603      &$ 17.0^{+10.7}_{-7.3}$  & $ 13.7^{+ 9.4}_{-6.5}$ & $ 13.6^{+ 9.8}_{-6.6}$ &$ 12.3^{+10.3}_{-7.8}$\\
\\                                                                                        
{\em Perseus:} \\                                                                         
King~1        &$-33.8^{+ 1.1}_{-1.2}$  & $-31.3^{+ 1.0}_{-1.1}$ & $-30.5^{+ 1.0}_{-1.0}$ &$-7.1^{+11.6}_{-11.6}$\\
NGC~559       &$-36.3^{+ 1.3}_{-1.3}$  & $-33.3^{+ 1.1}_{-1.2}$ & $-32.3^{+ 1.1}_{-1.1}$ &$-25.1^{+ 6.9}_{-6.9}$\\
NGC~7245      &$-43.2^{+ 4.7}_{-5.5}$  & $-40.0^{+ 4.2}_{-5.0}$ & $-39.5^{+ 4.1}_{-4.8}$ &$-25.3^{+ 5.3}_{-5.9}$ \\
\\                                                                                        
{\em Outer:} \\                                                                            
Berkeley~23   &$ 34.3^{+ 0.6}_{-0.6}$  & $ 32.3^{+ 0.5}_{-0.5}$ & $ 31.2^{+ 0.5}_{-0.5}$ &$ 16.3^{+ 3.4}_{-3.4}$ \\
\hline
\end{tabular}

(1): $\Theta_0=240$ km s$^{-1}$, $R_0=8.34$ kpc; \cite{2014ApJ...783..130R} \\
(2): $\Theta_0=220$ km s$^{-1}$, $R_0=8.50$ kpc; \cite{2011MNRAS.418.1423A}\\
(3): $\Theta_0=200$ km s$^{-1}$, $R_0=8.00$ kpc; \cite{2009PASJ...61..227S}\\
$d\Theta/dR=-0.2$ km s$^{-1}$ kpc$^{-1}$;  \cite{2014ApJ...783..130R} \\
\end{table}

Mean radial velocities were combined with proper motions to derive full
spatial velocities. Mean proper motions were taken from \cite{2014A&A...564A..79D}
and are listed in Table~\ref{table:pm}. These proper motions are based on the
UCAC4 catalogue \citep{2013AJ....145...44Z}. A comparison between UCAC4
and PPMXL \citep{2010AJ....139.2440R} proper motions for the stars in common
in the entire cluster area yields large differences, as large as the proper motions
themselves. For our radial velocity members, there is no coherence among
proper motions, either in the UCAC4 or in the PPMXL, or between the UCAC4 and the PPMXL. 
In spite of these uncertainties, we computed the non-circular velocity 
components ($U_s$, $V_s$, $W_s$), which
are also listed in Table~\ref{table:pm}. To estimate the uncertainties, 
we performed a classical Markov chain Monte Carlo simulation for each cluster 
with 10\,000 random realizations, accounting for
the uncertainties in distance modulus, proper motions, and radial velocity.
The mean values of the 10\,000 cases are taken as the estimated 
velocity components and their
standard deviation as the estimation of the uncertainties. These uncertainties
reflect the large uncertainties in the proper motions and prevent us from deriving
kinematics in the arms.

\begin{table*}
\caption{Mean proper motions and their errors from \cite{2014A&A...564A..79D} 
for our cluster sample.  $U_s$, $V_s$, and $W_s$ are the components of the 
non-circular velocity at the position of each cluster. They are computed 
from proper motions and radial velocities using the values for the motion 
of the Sun with respect to the LSR from \cite{2014ApJ...783..130R} and 
Galactic rotation as $\Theta_0=220$ km s$^{-1}$, $R_0=8.50$ kpc, and 
$d\Theta/dR=-0.2$ km s$^{-1}$ kpc$^{-1}$.}
\label{table:pm}
\begin{tabular}{lccccc}
\hline
Cluster       & $\mu_\alpha$cos$\delta$ &$\mu_\delta$ &  $U_s$ & $V_s$ & $W_s$\\
             &(mas yr$^{-1}$) & (mas yr$^{-1}$) & (km s$^{-1}$) & (km s$^{-1}$)& (km s$^{-1}$)\\
\hline
{\em Scutum:} \\
NGC~6603      & 1.06$\pm$4.06 & $-0.74\pm$5.12 &   4.68$\pm$26.75&  22.11$\pm$ 72.15& $-13.46\pm$69.58 \\
\\                                                                                                   
{\em Perseus:} \\                                                                                     
King~1        &$-2.43\pm$2.59 &  1.02$\pm$1.50 &  34.91$\pm$44.01&  17.43$\pm$ 36.75&  21.55$\pm$15.70 \\
NGC~559       &$-4.58\pm$2.95 &  1.41$\pm$1.46 &  86.82$\pm$47.90&  58.09$\pm$ 52.77&   5.04$\pm$18.88 \\
NGC~7245      &$-1.98\pm$3.58 & $-1.76\pm$3.19 &  16.99$\pm$73.88& $-18.66\pm$ 45.73&  16.73$\pm$69.93 \\
\\                                                                                                   
{\em Outer:} \\                                                                                       
Berkeley~23   & 2.65$\pm$2.10 & $-7.32\pm$4.68 &   9.55$\pm$13.39&$-204.98\pm$116.58& $-10.75\pm$78.99 \\
\hline
\end{tabular}
\end{table*}

\section{Conclusions}\label{sec6}

We have obtained medium-resolution spectra ($R\sim$ 8000) in the near-infrared 
CaT region ($\sim$8\,500 \AA) for 64 stars in the line of sight towards five OCs: 
Berkeley~23, King~1, NGC~559, NGC~6603, and NGC~7245. To our knowledge, these 
are the first radial velocity and 
metallicity determinations from spectroscopy available in the literature, 
except for NGC~6603, where radial velocities of four members were previously measured by 
\citetads{2008AJ....136..118F}. Our main results may be summarized as follows:

\begin{itemize}
\item Of the 15 stars analysed in Berkeley~23, only eight were determined 
from their radial velocities to be members, with one of them a main sequence star. 
We derived a mean radial velocity of $\left\langle V_{r}\right\rangle = 48.6\pm3.4$ km s$^{-1}$
from these eight members. 
Excluding the main sequence star, we derived [Fe/H] $=-0.42\pm0.13$.
\item Ten stars were sampled in the King~1 area. However, they do not 
define a clear radial velocity distribution. Moreover, they do not trace a clear 
sequence in the magnitude-$\Sigma$Ca planes for any of the three luminosity indicators used.
In any case, we derived a mean radial velocity and metallicity for this 
cluster in the same way as for the others, with the uncertainties in these 
determinations accounting for the observed dispersion. 
We obtained an average radial velocity of $\langle V_{\rm r} 
\rangle=-38.4\pm11.6$ km s$^{-1}$ and a mean metallicity of [Fe/H] $=+0.01\pm0.52$.
These results should be treated with caution because we 
cannot ensure that we have sampled real cluster members. 
\item For NGC~559, we derived a mean radial velocity of $\langle V_{\rm r} \rangle=-58.4\pm6.8$  
km s$^{-1}$ from 15 of the 18 stars observed. The metallicity obtained is 
[Fe/H] $=-0.25\pm0.14$.
\item For NGC~6603, one of the closest known OCs to the Galactic centre, we 
determined an average radial velocity of $\langle V_{\rm r} \rangle=26.0\pm4.3$  km 
s$^{-1}$ from 7 of the 11 stars observed. The metallicity derived  
([Fe/H] $=+0.43\pm0.15$) places this cluster within the most metal-rich 
OCs known.
\item In the case of NGC~7245, the mean redial velocity, $\langle 
V_{\rm r} \rangle=-65.3\pm3.2$ km s$^{-1}$, was obtained from five of the ten stars observed.
We obtained an average metallicity of [Fe/H] $=-0.15\pm0.18$.
\end{itemize}

We compared the properties derived here for the five clusters in our sample 
with the trends observed for other OCs in the Galactic disc. All the 
clusters studied  follow the trend described by other coeval OCs located at 
similar Galactocentric distances very well. NGC~6603 may play an important role in our 
understanding of the trends observed in the Galactic disc. If the 
high metallicity derived here is confirmed by high-resolution spectroscopy, the 
metallicity gradient in the inner disc could be even steeper than 
previously thought. 

The mean radial velocities $\langle V_{\rm r} \rangle$ were used to derive
the peculiar radial velocity of the clusters with respect to their LSR
$V_{\rm r}^{\rm pec}$. These values are compatible with those of high-mass
star-forming regions in the same arms \citep{2014ApJ...783..130R}.
NGC~559 and NGC~7245 in the Perseus arm have the same age and the same
$V_{\rm r}^{\rm pec}$, and have compatible [Fe/H], too. These features, taken together, may
indicate some kind of relationship in their formation.

We also derived full spatial velocities for the clusters by combining 
our mean radial velocities with mean proper motions from the literature. 
However, the uncertainty in the proper motions prevents any further study
of the kinematics or any meaningful comparison with the velocities of the
high-mass star-forming regions mentioned above.

\begin{acknowledgements}
We acknowledge the anonymous referee for their 
useful comments which have contributed to clarify the presentation of our 
results. This research made use of the WEBDA database, operated at the Department 
of Theoretical Physics and Astrophysics of the Masaryk University, and the 
SIMBAD database, operated at the CDS, 
Strasbourg, France. This publication makes use of data products from the Two 
Micron All Sky Survey, which is a joint project of the University of 
Massachusetts and the Infrared Processing and Analysis Center/California 
Institute of Technology, funded by the National Aeronautics and Space 
Administration and the National Science Foundation.
This research was supported by the MINECO (Spain's Ministry of Economy and Competitiveness) - 
FEDER through grants AYA2013-42781P, AYA2012-39551-C02-01, AYA2010-16717, AYA2008-01839 and ESP2013-48318-C2-1-R.
\end{acknowledgements}


\Online
\setcounter{table}{1}

\onltab{
\begin{sidewaystable*}
\caption{Observing logs and program star information.}
\label{obsstars}
\centering
\renewcommand{\footnoterule}{}
\begin{scriptsize}
\begin{tabular}{l c c c c c c c c c c c c c c c c c c c}
\hline\hline
Cluster & Star\footnote{Identifications taken from WEBDA database, except for NGC~559, which came from \citetads{2014MNRAS.437..804J}.}  & $\alpha_{2000}$ & $\delta_{2000}$ &     $V$   &   $B$     &   $I$     &     $K_\mathrm{S}$    &     $V_{r}$      &   EW$_{8498}$   &   EW$_{8442}$  &    EW$_{8662}$  &  [Fe/H]$_V$       &  [Fe/H]$_I$      &  [Fe/H]$_K$      &  t$_{exp}$     &  S/N &  Date    &   Note \\
       &       &   (hrs)         &      (deg)      &   (mag) & (mag)   & (mag)   &   (mag)  &   (km s$^{-1}$)  &     (\AA)       &      (\AA)     &      (\AA)      &                 &           &                      &   (sec)        &  per pixel            &          &        \\ \hline
Be23    & W112  &  06:33:15.9     &  +20:31:04.1    &  14.977 & 16.368  &         &  11.736  &  50.84$\pm$3.95  &  1.76$\pm$0.05  &  3.64$\pm$0.07 &   2.41$\pm$0.08 &  -0.25$\pm$0.10 & 21.73$\pm$3.09 & -0.29$\pm$0.06 &  2x700        &  32             & 17012013 &    0   \\ 
       & W104  &  06:33:15.2     &  +20:32:05.8    &  14.968 & 16.275  &         &  11.886  &  46.82$\pm$3.44  &  1.65$\pm$0.05  &  3.83$\pm$0.09 &   2.27$\pm$0.06 &  -0.28$\pm$0.10 & 21.61$\pm$3.07 & -0.28$\pm$0.06 &  2x700        &  32             & 17012013 &    0   \\ 
       & W124  &  06:33:18.4     &  +20:32:06.6    &  15.268 & 16.467  &         &  12.394  &  47.65$\pm$4.05  &  1.20$\pm$0.07  &  3.35$\pm$0.08 &   2.27$\pm$0.06 &  -0.58$\pm$0.10 & 19.93$\pm$2.90 & -0.59$\pm$0.06 &  2x700        &  29             & 17012013 &    0   \\ 
       & W159  &  06:33:19.9     &  +20:29:28.0    &  16.166 & 16.813  &         &        &  51.47$\pm$10.91 &                 &                &                 &                 &             &                &  2x700 &   9          & 17012013 &    2   \\ 
       & W089  &  06:33:22.3     &  +20:32:51.9    &  14.808 & 16.064  &         &  11.866  &  54.92$\pm$3.25  &  1.30$\pm$0.05  &  3.19$\pm$0.06 &   2.87$\pm$0.07 &  -0.49$\pm$0.10 & 20.90$\pm$3.00 & -0.46$\pm$0.06 &  2x700        &  31             & 17012013 &    0   \\ 
       & W079  &  06:33:09.2     &  +20:27:21.3    &  14.460 & 15.865  &         &  11.090  &  46.39$\pm$4.76  &  1.45$\pm$0.05  &  3.54$\pm$0.06 &   2.65$\pm$0.05 &  -0.48$\pm$0.10 & 21.40$\pm$3.05 & -0.51$\pm$0.05 &  2x650       &  39              & 18012013 &    0   \\ 
       & W092  &  06:33:00.2     &  +20:31:44.1    &  14.702 & 16.095  &         &  11.476  &  44.89$\pm$4.64  &  1.53$\pm$0.07  &  3.30$\pm$0.08 &   2.56$\pm$0.09 &  -0.51$\pm$0.10 & 20.94$\pm$3.00 & -0.54$\pm$0.06 &  2x650   &  26          & 18012013 &    0   \\ 
       & W106  &  06:33:18.4     &  +20:33:31.6    &  15.072 & 16.309  &         &  12.191  &  46.24$\pm$5.20  &  1.44$\pm$0.09  &  3.78$\pm$0.07 &   2.38$\pm$0.08 &  -0.31$\pm$0.10 & 21.32$\pm$3.04 & -0.29$\pm$0.06 &  2x700   &  23          & 18012013 &    0   \\ 
       & W135  &  06:33:17.0     &  +20:32:06.3    &  16.011 & 16.551  &         &  14.347  &  99.82$\pm$11.92 &                 &                &                 &                 &           &                &  2x700         &  17   & 17012013 &    1   \\ 
       & W102  &  06:33:19.2     &  +20:31:34.5    &  14.881 & 16.262  &         &  11.666  &  26.07$\pm$3.11  &                 &                &                 &                 &           &                &  2x700         &  34   & 17012013 &    1   \\ 
       & W141  &  06:33:20.1     &  +20:31:17.9    &  15.383 & 16.607  &         &  12.506  &  26.33$\pm$4.47  &                 &                &                 &                 &           &                &  2x700         &  25   & 17012013 &    1   \\ 
       & W091  &  06:33:21.8     &  +20:29:49.8    &  14.747 & 16.081  &         &  11.724  &   4.96$\pm$3.51  &                 &                &                 &                 &           &                &  2x700         &  24   & 17012013 &    1   \\ 
       & W025  &  06:33:17.6     &  +20:26:06.5    &  13.832 & 14.505  &         &  12.019  & -17.84$\pm$3.86  &                 &                &                 &                 &           &                &  2x700         &  22   & 17012013 &    1   \\
       & W038  &  06:33:20.6     &  +20:26:23.5    &  13.557 & 15.226  &         &   9.911  & -85.37$\pm$5.22  &                 &                &                 &                 &           &                &  2x700         &  67   & 17012013 &    1   \\ 
       & W059  &  06:33:01.9     &  +20:35:33.5    &  14.052 & 15.526  &         &  10.720  &  62.36$\pm$5.14  &                 &                &                 &                 &           &                &  2x600         &  41   & 18012013 &    1   \\ 
King1   & W0117 &  00:21:50.8     &  +64:30:13.6    &  14.062 & 15.712  & 11.901  &   9.773  & -26.87$\pm$2.03  &  1.49$\pm$0.05  &  4.32$\pm$0.04 &   3.23$\pm$0.05 & 0.36$\pm$0.08 & 0.26$\pm$0.06 & 0.12$\pm$0.04 &  2x300        &  44   & 19072014 &    3   \\ 
       & W0296 &  00:22:47.5     &  +64:28:54.1    &  12.985 & 14.894  & 10.451  &   7.874  & -46.94$\pm$1.98  &  1.97$\pm$0.03  &  5.32$\pm$0.03 &   4.64$\pm$0.04 &   1.16$\pm$0.09 &  1.04$\pm$0.06 &  0.89$\pm$0.04 &  2x150   &  60          & 19072014 &    3   \\ 
       & W0405 &  00:22:43.5     &  +64:28:18.8    &  13.901 & 15.612  & 11.524  &   9.240  & -31.64$\pm$2.34  &  1.18$\pm$0.04  &  3.43$\pm$0.05 &   2.52$\pm$0.05 &  -0.49$\pm$0.07 & -0.66$\pm$0.05 & -0.85$\pm$0.04 &  2x300   &  46          & 19072014 &    3   \\ 
       & W0971 &  00:22:22.5     &  +64:25:07.9    &  14.183 & 15.869  & 11.979  &   9.777  & -35.58$\pm$3.01  &  1.72$\pm$0.04  &  4.28$\pm$0.04 &   3.13$\pm$0.04 &   0.44$\pm$0.08 &  0.32$\pm$0.06 &  0.16$\pm$0.04 &  2x300 &  43            & 19072014 &    3   \\ 
       & W1092 &  00:22:18.1     &  +64:24:39.1    &  13.327 & 15.17    & 10.885  &   8.398  & -34.13$\pm$1.51  &  1.63$\pm$0.03  &  4.44$\pm$0.03 &   3.11$\pm$0.04 &   0.17$\pm$0.08 &  0.04$\pm$0.05 & -0.15$\pm$0.04 &  2x250   &  62          & 19072014 &    3   \\ 
       & W1280 &  00:22:09.0     &  +64:23:46.8    &  13.756 & 15.759  & 11.141  &   8.488  & -37.36$\pm$1.88  &  1.45$\pm$0.03  &  4.24$\pm$0.03 &   3.17$\pm$0.03 &   0.19$\pm$0.07 & -0.02$\pm$0.05 & -0.27$\pm$0.04 &  2x250   &  65          & 19072014 &    3   \\ 
       & W1775 &  00:21:50.7     &  +64:21:31.5    &  14.052 & 15.835  & 11.742  &   9.477  & -51.40$\pm$2.23  &  0.75$\pm$0.04  &  3.18$\pm$0.06 &   2.46$\pm$0.09 &  -0.76$\pm$0.08 & -0.92$\pm$0.06 & -1.13$\pm$0.05 &  2x300   &  47          & 19072014 &    3   \\ 
       & W2104 &  00:22:34.2     &  +64:19:44.9    &  13.593 & 15.421  & 11.135  &   8.705  & -56.65$\pm$2.42  &  1.76$\pm$0.03  &  4.45$\pm$0.02 &   3.04$\pm$0.04 &   0.29$\pm$0.08 &  0.14$\pm$0.05 & -0.05$\pm$0.04 &  2x300   &  58          & 19072014 &    3   \\ 
       & W2282 &  00:21:35.5     &  +64:19:03.1    &  14.139 & 15.867  & 11.874  &   9.565  & -45.21$\pm$1.81  &  1.39$\pm$0.04  &  3.95$\pm$0.05 &   2.86$\pm$0.04 &   0.03$\pm$0.08 & -0.11$\pm$0.06 & -0.31$\pm$0.04 &  2x300   &  50          & 19072014 &    3   \\ 
       & W2410 &  00:21:27.2     &  +64:18:27.4    &  14.049 & 15.826  & 11.815  &   9.512  & -18.53$\pm$1.88  &  1.46$\pm$0.04  &  4.22$\pm$0.04 &   2.76$\pm$0.04 &   0.11$\pm$0.08 & -0.02$\pm$0.05 & -0.21$\pm$0.04 &  2x300   &  41          & 19072014 &    3   \\ 
NGC559  & J0010 &  01:29:36.7     &  +63:22:08.0    &  12.792 & 14.927  & 10.065  &   7.215  & -11.13$\pm$1.70  &                 &                &                 &                 &           &      &  2x300     &  80              & 18072014 &    1   \\ 
       & J0018 &  01:29:31.1     &  +63:18:12.5    &  13.442 & 15.061  & 11.514  &   9.400  & -65.48$\pm$2.51  &  1.83$\pm$0.05  &  4.21$\pm$0.05 &   2.88$\pm$0.05 &  -0.10$\pm$0.08 & -0.01$\pm$0.06 & -0.17$\pm$0.04 &  2x300   &  50          & 18072014 &    0   \\ 
       & J0020 &  01:29:46.0     &  +63:19:27.0    &  13.529 & 15.350  & 11.306  &   8.861  & -61.91$\pm$1.62  &  1.44$\pm$0.04  &  4.42$\pm$0.03 &   2.78$\pm$0.04 &  -0.18$\pm$0.07 & -0.19$\pm$0.05 & -0.41$\pm$0.04 &  2x300   &  50          & 18072014 &    0   \\ 
       & J0025 &  01:28:41.2     &  +63:20:47.8    &  13.612 & 15.100  & 11.700  &   9.716  & -63.63$\pm$3.14  &  1.21$\pm$0.04  &  4.00$\pm$0.04 &   3.00$\pm$0.03 &  -0.33$\pm$0.07 & -0.26$\pm$0.05 & -0.40$\pm$0.04 &  2x300   &  46          & 19072014 &    0   \\ 
       & J0027 &  01:29:45.6     &  +63:18:42.2    &  13.716 & 15.218  & 11.864  &   9.868  & -55.26$\pm$2.07  &  1.36$\pm$0.04  &  3.79$\pm$0.05 &   2.65$\pm$0.05 &  -0.46$\pm$0.07 & -0.39$\pm$0.05 & -0.54$\pm$0.04 &  2x300   &  43          & 18072014 &    0   \\ 
       & J0030 &  01:29:20.9     &  +63:17:36.4    &  13.758 & 15.372  & 11.800  &   9.667  & -51.97$\pm$2.54  &  1.55$\pm$0.06  &  4.29$\pm$0.09 &   3.11$\pm$0.07 &   0.01$\pm$0.09 &  0.07$\pm$0.07 & -0.09$\pm$0.06 &  2x300   &  32          & 18072014 &    0   \\ 
       & J0033 &  01:30:08.0     &  +63:18:33.8    &  13.781 & 15.256  & 11.999  &  10.110  & -57.90$\pm$2.54  &  1.38$\pm$0.05  &  4.35$\pm$0.10 &   2.81$\pm$0.06 &  -0.14$\pm$0.09 & -0.04$\pm$0.07 & -0.16$\pm$0.06 &  2x300   &  33          & 18072014 &    0   \\ 
       & J0034 &  01:29:59.1     &  +63:22:37.8    &  13.794 & 15.380  & 11.780  &   9.687  & -70.71$\pm$2.24  &  1.49$\pm$0.04  &  4.29$\pm$0.06 &   2.89$\pm$0.05 &  -0.08$\pm$0.08 & -0.05$\pm$0.06 & -0.21$\pm$0.05 &  2x300   &  49          & 18072014 &    0   \\ 
       & J0038 &  01:29:29.3     &  +63:18:05.6    &  13.937 & 15.578  & 11.990  &   9.792  & -66.61$\pm$1.85  &  1.51$\pm$0.03  &  3.94$\pm$0.03 &   2.67$\pm$0.04 &  -0.26$\pm$0.07 & -0.23$\pm$0.05 & -0.42$\pm$0.04 &  2x300   &  62          & 18072014 &    0   \\ 
       & J0040 &  01:29:41.5     &  +63:17:55.1    &  13.965 & 15.544  & 11.987  &   9.876  & -59.81$\pm$2.06  &  1.57$\pm$0.02  &  4.08$\pm$0.04 &   2.71$\pm$0.06 &  -0.16$\pm$0.08 & -0.13$\pm$0.06 & -0.30$\pm$0.04 &  2x300   &  68          & 19072014 &    0   \\ 
       & J0041 &  01:29:58.2     &  +63:18:21.1    &  13.972 & 15.518  & 12.058  &   9.991  &   5.16$\pm$2.47  &                 &                &                 &                 &           &                &  2x300       &  42            & 18072014 &    1   \\ 
       & J0043 &  01:29:32.2     &  +63:18:45.7    &  14.000 & 15.517  & 12.110  &  10.060  & -59.41$\pm$2.45  &  1.42$\pm$0.03  &  3.86$\pm$0.04 &   2.73$\pm$0.04 &  -0.29$\pm$0.07 & -0.24$\pm$0.05 & -0.41$\pm$0.04 &  2x300   &  57          & 19072014 &    0   \\ 
       & J0044 &  01:29:34.2     &  +63:19:19.6    &  14.001 & 15.730  & 11.875  &   9.575  & -55.86$\pm$2.32  &  1.41$\pm$0.02  &  3.94$\pm$0.03 &   3.09$\pm$0.03 &  -0.11$\pm$0.07 & -0.12$\pm$0.05 & -0.34$\pm$0.04 &  2x300   &  79          & 18072014 &    0   \\ 
       & J0049 &  01:29:42.0     &  +63:16:52.4    &  14.049 & 15.784  & 11.914  &   9.575  & -57.81$\pm$2.04  &  1.62$\pm$0.03  &  4.15$\pm$0.03 &   2.76$\pm$0.04 &  -0.06$\pm$0.07 & -0.07$\pm$0.05 & -0.29$\pm$0.04 &  2x300   &  57          & 19072014 &    0   \\ 
       & J0061 &  01:29:15.3     &  +63:18:07.3    &  14.236 & 15.728  & 12.411  &  10.427  & -51.09$\pm$2.13  &  1.43$\pm$0.04  &  3.46$\pm$0.05 &   2.71$\pm$0.04 &  -0.38$\pm$0.07 & -0.34$\pm$0.05 & -0.51$\pm$0.04 &  2x300   &  46          & 18072014 &    0   \\ 
       & J0067 &  01:29:34.2     &  +63:20:29.9    &  14.307 & 15.810  & 12.434  &  10.410  & -54.68$\pm$2.95  &  1.32$\pm$0.06  &  3.57$\pm$0.06 &   2.60$\pm$0.08 &  -0.41$\pm$0.08 & -0.38$\pm$0.06 & -0.56$\pm$0.05 &  2x300   &  32          & 18072014 &    0   \\ 
       & J0070 &  01:30:04.8     &  +63:19:26.3    &  14.380 & 15.884  & 12.490  &  10.494  & -37.12$\pm$2.72  &                 &                &                 &                    &                &                &  2x300   &  32            & 18072014 &    1   \\ 
       & J0082 &  01:29:48.6     &  +63:15:29.1    &  14.527 & 16.128  & 12.525  &  10.286  & -44.17$\pm$4.55  &  1.42$\pm$0.03  &  3.56$\pm$0.05 &   2.63$\pm$0.05 &  -0.29$\pm$0.07 & -0.31$\pm$0.06 & -0.54$\pm$3.05 &  2x300     &  36       & 18072014 &    0   \\ 
NGC6603 & W1653 &  18:18:31.5     &  -18:25:48.0    &  13.62  & 15.42   & 11.72   &   8.976  & -32.44$\pm$1.78  &                 &                &                 &                 &           &                &  2x400         &  86   & 18072014 &    1   \\ 
        & W1727 &  18:18:30.7     &  -18:24:24.4    &  13.29  & 14.78   & 11.72     &   9.351  &  -5.40$\pm$2.42  &                 &                &                 &                 &         &                &  2x350         &  62   & 19072014 &    1   \\ 
        & W1755 &  18:18:30.5     &  -18:25:04.0    &  13.68  & 15.52   & 11.86     &   8.884  &  63.55$\pm$2.10  &                 &                &                 &                 &         &                &  2x400         &  97   & 18072014 &    1   \\ 
        & W1997 &  18:18:28.5     &  -18:24:57.0    &  13.61  & 15.28   & 11.81     &   9.346  &  23.58$\pm$2.38  &  1.77$\pm$0.02  &  4.96$\pm$0.02 &   4.19$\pm$0.03 &  0.60$\pm$0.24  &  0.73$\pm$0.19 &  0.57$\pm$0.15 &  2x400 &  82            & 18072014 &    0   \\ 
        & W2033 &  18:18:28.1     &  -18:24:35.0    &  13.76  & 15.52   & 11.96     &   9.449  &  28.50$\pm$2.89  &  1.66$\pm$0.03  &  4.90$\pm$0.02 &   3.59$\pm$0.03 &  0.36$\pm$0.23  &  0.46$\pm$0.18 &  0.27$\pm$0.14 &  2x400   &  79          & 18072014 &    0   \\ 
        & W2215 &  18:18:26.3     &  -18:24:00.0    &  13.40  & 15.17   & 11.63     &   9.039  &  27.48$\pm$2.40  &  1.57$\pm$0.02  &  4.94$\pm$0.02 &   4.18$\pm$0.03 &  0.44$\pm$0.24  &  0.59$\pm$0.18 &  0.34$\pm$0.15 &  2x400 &  88            & 18072014 &    0   \\ 
        & W2249 &  18:18:26.2     &  -18:25:37.6    &  13.20  & 14.67   & 11.53     &   9.228  &  33.04$\pm$4.14  &  1.80$\pm$0.02  &  4.77$\pm$0.03 &   3.68$\pm$0.05 &  0.20$\pm$0.24  &  0.38$\pm$0.18 &  0.26$\pm$0.14 &  2x350s  &  88          & 19072014 &    0   \\ 
        & W2252 &  18:18:25.9     &  -18:23:54.0    &  13.63  & 15.43   & 11.78     &   9.155  &  25.83$\pm$1.68  &  1.51$\pm$0.03  &  5.22$\pm$0.02 &   3.58$\pm$0.05 &  0.37$\pm$0.24  &  0.47$\pm$0.18 &  0.26$\pm$0.15 &  2x400   &  75          & 18072014 &    0   \\ 
        & W2352 &  18:18:25.2     &  -18:25:01.0    &  13.65  & 15.17   & 11.89     &   9.403  &  24.44$\pm$2.42  &  1.60$\pm$0.03  &  4.51$\pm$0.02 &   4.60$\pm$0.03 &  0.54$\pm$0.24  &  0.67$\pm$0.18 &  0.50$\pm$0.15 &  2x400   &  68          & 19072014 &    0   \\ 
        & W2438 &  18:18:24.5     &  -18:25:44.4    &  13.17  & 15.4    & 10.75     &   7.469  &  19.32$\pm$3.10  &  1.87$\pm$0.02  &  5.24$\pm$0.01 &   3.78$\pm$0.02 &  0.43$\pm$0.24  &  0.40$\pm$0.19 &  0.07$\pm$0.15 &  2x400   & 118          & 19072014 &    0   \\ 
        & W2492 &  18:18:23.9     &  -18:26:19.0    &  13.49  & 15.15   & 11.73     &   9.079  & -50.21$\pm$3.74  &                 &                &                 &          &                &                &  2x400         &  69   & 19072014 &    1   \\ 
NGC7245 & W0045 &  22:15:07.8     &  +54:18:26.9    &  14.165 & 15.521  & 12.712  &  10.873  & -39.48$\pm$2.65  &                 &                &                 &                 &           &                &  2x200         &  27   & 19072014 &    1   \\ 
       & W0055 &  22:15:17.5     &  +54:18:12.6    &  13.109 & 14.385  & 11.738  &   9.952  & -69.42$\pm$2.78  &  1.67$\pm$0.04  &  4.66$\pm$0.04 &   3.01$\pm$0.04 &   0.06$\pm$0.11 &  0.13$\pm$0.10 &  0.20$\pm$0.10 &  2x150   &  40          & 19072014 &    0   \\ 
       & W0178 &  22:15:05.4     &  +54:22:43.6    &  13.758 & 15.006  & 12.392  &   9.711  & -66.92$\pm$3.97  &  1.67$\pm$0.03  &  3.93$\pm$0.04 &   2.46$\pm$0.04 &  -0.24$\pm$0.10 & -0.23$\pm$0.10 & -0.42$\pm$0.10 &  2x120   &  28          & 18072014 &    0   \\ 
       & W0179 &  22:15:05.4     &  +54:22:49.4    &  12.969 & 14.329  & 11.574  &   9.711  & -60.81$\pm$4.33  &  1.58$\pm$0.06  &  4.30$\pm$0.07 &   3.02$\pm$0.06 &  -0.16$\pm$0.11 & -0.10$\pm$0.11 & -0.05$\pm$0.11 &  2x120   &  37          & 18072014 &    0   \\ 
       & W0205 &  22:15:14.9     &  +54:20:04.1    &  13.866 & 15.203  & 12.418  &  10.550  & -64.73$\pm$2.38  &  1.50$\pm$0.06  &  3.69$\pm$0.07 &   2.67$\pm$0.07 &  -0.29$\pm$0.10 & -0.31$\pm$0.10 & -0.32$\pm$0.10 &  2x220   &  38          & 18072014 &    0   \\ 
       & W3154 &  22:14:56.3     &  +54:21:51.6    &  13.459 & 14.831  & 12.037  &  10.136  & -27.63$\pm$1.73  &                 &                &                 &                 &           &                &  2x200         &  45   & 19072014 &    1   \\ 
       & W3214 &  22:14:58.9     &  +54:23:05.4    &  13.618 & 14.865  & 12.276  &  10.537  & -51.73$\pm$2.31  &                 &                &                 &            &                &                &  2x220         &  38   & 18072014 &    1   \\ 
       & W3391 &  22:15:09.5     &  +54:23:52.4    &  14.295 & 15.639  & 12.852  &  10.932  & -12.13$\pm$4.55  &                 &                &                 &            &                &                &  2x200+2x150 &  33    & 18072014 &    1   \\ 
       & W0073 &  22:15:04.4     &  +54:20:18.7    &  13.265 & 14.478  & 11.979  &  10.328  & -18.05$\pm$1.66  &                 &                &                 &                 &           &                &  2x220         &  54   & 18072014 &    1   \\ 
       & W0095 &  22:15:12.0     &  +54:21:11.4    &  13.373 & 14.666  & 12.002  &  10.211  & -64.68$\pm$3.06  &  1.98$\pm$0.06  &  4.27$\pm$0.05 &   2.17$\pm$0.08 &  -0.21$\pm$0.10 & -0.18$\pm$0.10 & -0.15$\pm$0.10 &  2x220        &  36             & 18072014 &    0   \\  \hline\hline 
\end{tabular}\\
\tablefoot{(0) Cluster member from radial velocity (RGB or RC); (1) cluster non-member from radial velocity; (2) Cluster member from radial velocity (main sequence); (3) Unclear membership.}
\end{scriptsize}
\end{sidewaystable*}

}

\end{document}